\documentclass[preprint,aps]{revtex4}
\begin{document}
\title {Josephson current in ballistic Nb/InAs/Nb highly transmissive junctions}
\author{Francesco Giazotto}
\email{giazotto@sns.it}
\author{Kasper Grove-Rasmussen }
\author{Rosario Fazio }
\author{Fabio Beltram}
\affiliation{NEST-INFM \& Scuola Normale Superiore, I-56100 Pisa, Italy}

\author{Edmund H. Linfield and David A. Ritchie}
\affiliation{Cavendish Laboratory, University of Cambridge, Madingley Road, Cambridge CB3 0HE, United Kingdom}

\begin{abstract}
Highly transmissive ballistic junctions are demonstrated between Nb and the two-dimensional electron gas formed at an InAs/AlSb heterojunction.
A reproducible fabrication protocol is presented yielding high critical supercurrent values. Current-voltage characteristics were measured down to 0.4 K and the observed supercurrent behavior was analyzed within a ballistic model in the clean limit. This investigation allows us to demonstrate an intrinsic interface transmissivity approaching 90\%. The reproducibility of the fabrication protocol makes it of interest for the experimental study of InAs-based superconductor-semiconductor hybrid devices. 
\end{abstract}


\maketitle

Superconductor-semiconductor-superconductor (S-Sm-S) hybrid  devices have been of considerable interest over recent years as a result of their potential for use in a significant number of applications \cite{ruggero}. 
Amongst these the Josephson field effect transistor (JoFET) \cite{tak} is of particular technological relevance. Operation of the JoFET is based on the modulation of the Josephson coupling between two adjacent S contacts by the variation of the carrier density of the Sm layer separating them. 
In these hybrid structures supercurrent originates from current-carrying bound states \cite{kulik} due to \emph{Andreev reflection} \cite{andr}. The latter converts quasiparticle current in the semiconductor into Cooper-pair current in the superconductor.
Key issues to be addressed in order to make these devices of practical interest include high S-Sm interface transmissivity (as required for an efficient Andreev reflection) and fabrication schemes allowing inter-electrode distances ($L$) much smaller than the semiconductor coherence length $\xi_{Sm}$ (in fact critical supercurrent values decrease proportionally to $e^{-L/\xi_{Sm}}$ \cite{likharev}). 
Favorable $L/\xi_{Sm}$ ratios can be achieved by electron beam lithography (EBL), yielding a small electrode spacing, in combination with Sm layers consisting of high-quality two-dimensional electron gases (2DEGs). The latter provide electron mean free path $\ell_m$, mobility and coherence-length values significantly exceeding those obtained in bulk semiconductors.
A more challenging issue is the fabrication of high-transmissivity S-Sm contacts. 
For GaAs-based 2DEGs the dominant factor limiting interface-transmissivity is represented by the presence of the Schottky barrier for which penetrating alloyed superconducting ohmic contacts proved to be effective in achieving good interface transparency \cite{gaas}. 
This method, however, suffers from the poor geometric definition intrinsic to the alloying process making it poorly compatible with high-resolution EBL. 
An efficient alternative is based on the exploitation of III-V semiconductor alloys with high In content \cite{ingaas} and in particular on InAs-based 2DEGs \cite{inas}. These can provide Schottky barrier-free metal-semiconductor contacts. The residual transparency-limiting factors are much less severe and stem from Fermi-velocity mismatch \cite{blonder} and/or interface contaminants.

In this letter we report the fabrication and characterization of highly transmissive Nb/2DEG/Nb ballistic junctions made in the InAs/AlSb system. 
Current-voltage characteristics were measured down to 0.4 K and the observed supercurrent behavior was analyzed within a ballistic model in the clean limit. This investigation allows us to demonstrate an intrinsic interface transmissivity close to 90\% in our system. The reproducibility of the fabrication protocol makes it of interest for the experimental study of InAs-based S-Sm hybrid devices. 

The semiconductor portion of the  structure (sketched in Fig. 1(a)) was grown by molecular beam epitaxy without intentional doping on a semi-insulating GaAs(100) substrate. The Sm link consists of a 10-nm-wide InAs quantum well sandwiched between AlSb barriers. A 5 nm GaSb cap layer was grown on top of the structure in order to protect the AlSb layer from oxidation.
%The growth sequence starts with a 0.5 $\mu$m GaAs layer followed by a buffer layer consisting of ten-period (25 \AA  + 25 \AA) AlSb/GaSb superlattice (SL) sandwiched between two AlSb buffer layers, 50 nm and 1 $\mu$m thick. The quantum well itself consists of a 10 nm InAs well sandwiched between a ten-period (25 \AA + 25 \AA) AlSb/GaSb SL barrier on the bottom, and a 20 nm-thick AlSb barrier on the top. Then, a 5 nm GaSb cap layer was grown on top of the structure in order to protect the AlSb from oxidation.
Standard photolithographic techniques and wet etching \cite{etching} were used to define a 20-$\mu$m-wide Hall-bar mesa. 
At $T=0.3$ K we measured a sheet electron concentration  $n\simeq 7.4\times 10^{11}$ cm$^{-2}$ and an electron mobility $\mu \simeq 75000$ cm$^2$/Vs. An effective mass $m^{\ast}=0.036\,\,m_e$, with $m_e$ the free electron mass, was deduced from temperature-dependent Shubnikov-de Haas measurements. Another useful parameter, the single-particle dephasing length $\ell _{\phi}$ was measured at $T=0.3$ K by weak localization magnetoresistance measurements and 
a value of 2.7 $\mu$m was obtained . These results allow us to calculate the  electron mean free path $\ell _m \simeq 1.1$ $\mu$m and a thermal coherence length in the clean limit $\xi_{Sm}(T)=\hbar v_F /2\pi k_B T =0.85$ $\mu$m/$T$, where $v_F$ is the Fermi velocity in the InAs layer. For $T\geq 0.8$ K, $\xi_{Sm}(T)<\ell_m$, i.e. electron transport in the sample is in the clean limit.

S/2DEG/S junctions were patterned with a single EBL step. First, two 20-$\mu$m-wide openings separated by a 190-nm-wide channel were defined in a single PMMA layer (see Fig. 1(b),(c)) and the GaSb/AlSb top layers were removed by wet etching in order to expose the InAs well. 
The sample was then loaded into an UHV deposition chamber (background pressure of $p=1.2 \times 10^{-10}$ Torr) to carry out the S electrode deposition. Before performing the latter, the exposed areas were subjected to RF sputter-cleaning procedure \cite{inas}. The cleaning treatment was performed at $p=12$ mTorr of Ar partial pressure, exposing the etched areas for 6 minutes to a 0.3 W/cm$^2$ plasma power density. 
Immediately after this, 50-nm-thick Nb electrodes were deposited {\it in situ} by DC-magnetron sputtering at a rate of  about 30 \AA/s.
The resulting Nb layer shows a critical temperature $T_c = 8.5$ K, corresponding to an energy gap $\Delta =1.4$ meV.
The Nb/InAs-2DEG/Nb junctions were electrically characterized in a closed-cycle $^3$He refrigerator from 0.4 K to temperatures larger than $T_c$ and current-driven four-terminal measurements were performed between the two Nb electrodes. 
 
Figure 2 shows the current-voltage characteristics of a typical Nb/InAs/Nb junction in the $0.4 \div 5.0$ K temperature range.
The curves display a well-developed supercurrent $I_c$ and no evidence of hysteresis was observed, as expected for overdamped junctions \cite{likharev}.
At  $T=0.4$ K  a 11 $\mu$A critical supercurrent was measured corresponding to a critical-current linear density of 0.55 A/m. To the best of our knowledge this value is among the highest reported for S/InAs-2DEG/S of comparable length and free-carrier concentration \cite{inas}.
Furthermore, from the junction normal-state resistance $R_N =15\,\,\Omega$, a characteristic product $I_c R_N =165\,\,\mu$V is extracted which suggests good interface morphology \cite{inas}. The inset of Fig. 2 displays the current-voltage characteristic at $T=0.4$ K over a wider bias range. The linear extrapolation of the curve from $eV\gg 2\Delta$ to zero bias (dashed line in the inset) allows us to determine the junction excess current, $I_{exc}\simeq 57$ $\mu$A. This quantity is an important figure of merit in S-Sm-S weak links
the magnitude of which strongly depends on interface quality. The excellence of this value can be understood by following the idealized one-dimensional S-Normal metal-S model in Ref. \cite{otbk} which yields a rather high intrinsic barrier transparency $Z \approx$ 0.65 \cite{nota}.

More information about the success of the fabrication protocol and on the properties of the hybrid system realized can be obtained by analyzing $I_c$ as a function of temperature in the temperature range up to $T_c$. In fact this analysis allows us to determine some crucial junction parameters like the true interface transmissivity.  
Figure 3(a) (full diamonds) shows the measured supercurrents normalized to the $I_c$ value at $T=0.4$ K as a function of the reduced temperature ($T/T_c$). 
The $I_c(T)$ behavior follows a characteristic trend and $I_c$s decrease with increasing temperatures \cite{likharev}. For temperatures lower than $T/T_c \approx 0.1$ data appear to saturate as expected for a non-perfectly transmissive S-Sm interface \cite{schussler,chrestin}. A quantitative description of the data can be performed, but much care must be taken in carrying out this analysis owing to the two-dimensional nature of the InAs region, which is coupled to bulk Nb electrodes.
Indeed supercurrent values may be deeply affected by the abrupt dimensionality change at the S-Sm contact, by the change in electron effective mass and by band-edge discontinuities. 

As mentioned above, the electrode separation ($L=190$ nm) favorably compares with $\ell_{m}$ ($\ell_{m}\gg L$) and allows a ballistic analysis of the system. This makes it possible to take advantage of the model developed by Chrestin and co-workers \cite{chrestin}. These authors considered the S/2DEG/S structure sketched in the inset of Fig. 3(a). 
It consists of an InAs-2DEG of length $L$ laterally contacted through identical potential barriers to two superconducting bulk leads. In our case of InAs-based weak links the contact transmissivity is limited by the Fermi-velocity mismatch between the S and the Sm layers and by contaminants at the interface resulting from the fabrication procedure. The influence of the latter on interface transparency can be conveniently modeled by a $\delta$ function weighted by a dimensionless parameter $Z$ \cite{btk}.   
The normal-state contact transmissivity $\mathcal T$ can be readily obtained from these two contributions as $\mathcal T = 4 \mathcal R [4 \mathcal R Z^2 + (1 + \mathcal R)^2]^{-1}$ \cite{blonder}, where $\mathcal R$ is the ratio of the Fermi velocities between S and Sm.
Solving the Bogolubov-de Gennes equations, the Josephson current in the system can be calculated as a function of the structure parameters. The temperature dependence of the critical current $I_c(T)$ can also be obtained and compared with our experiment.
Figure 3(a) (open circles) shows the best fit to the data computed with the model described above. 
It was calculated employing the parameters for InAs and Nb determined experimentally with a single fitting parameter: a barrier strength $Z=0.4$ \cite{parametri}. 
The experimental data are well described by this model over the whole temperature range thus further supporting our description of the system in the clean limit. 
The fitted value $Z=0.4$ translates into an intrinsic barrier transmissivity exceeding 86\% and in a total contact transmissivity of the order of 80\% (including the existing Nb/InAs Fermi-velocity mismatch). This very large value is consistent with the differential-conductance versus bias ($G(V)$) behavior in the $4.6\div 8.5$ K temperature range (see Fig. 3(b)). (In this temperature range the supercurrent was not measurable.) 
The curves show a large enhancement for energies lower than $2\Delta /e$ and a peculiar reproducible structure is  visible within this energy range on the $G(V)$ spectra. The non monotonic behavior observed can be ascribed to multiple Andreev reflections  off the Nb electrodes \cite{otbk}. The resulting resonances are too weak to allow a detailed comparison with the theory, but this very fact is fully consistent with the highly transmissive nature of S-Sm contacts as theoretically predicted in Ref. \cite{otbk}.

In summary, Nb/InAs-2DEG/Nb ballistic weak links were fabricated and characterized as a function of temperature. 
The devices shows a supercurrent of 11 $\mu$A at $T=0.4$ K and a good value for the product $I_c R_N=165$ $\mu$V.
A theoretical model in the clean limit describes well the experimental behavior of these structures and allows us to estimate a high value for the intrinsic contact transmissivity approaching 90\%. Our results were made possible by the fabrication procedure adopted here and suggest that the Nb/InAs-2DEG combination may be considered as the prototype system on which to implement successfully high-performance Josephson-type devices. 
   
The authors would like to acknowledge P. Pingue for fruitful discussions.
This work was supported by INFM under the PAIS projects EISS and TIN, and by EPSRC (UK). EHL acknowledges support from Toshiba Research Europe Ltd.

%------------------------------------------ References

%------ Figures beginning -------------

\begin{figure}
\caption{ (a) Sketch of the Nb/InAs-2DEG/Nb microstructure. The Nb-electrode separation is $L=190$ nm. (b) Scanning electron micrograph of the device. The mesa lateral arms were used as additional probes for electrical characterization. (c) 
Magnified view of Fig. 1(b) showing the semiconductor channel separating the two Nb contacts.}
\label{F1}
\end{figure}

\begin{figure}
\caption{Current-voltage characteristics of the Nb/InAs/Nb weak link in the $0.4\div 5$ K temperature range. 
Curves are horizontally offset for clarity. From left to right, data were taken at $T=0.4,\,1.0,\,1.9,\,2.0,\,2.5,\,2.8,\,3.1,\,3.6,\,4.0,\,4.5,$ and $5.0$ K.
The inset shows the current-voltage characteristic measured at $T=0.4$ K over a wider bias range. 
The linear extrapolation to $V=0$ yields an excess current of 57 $\mu$A. 
}
\label{F2}
\end{figure}

\begin{figure}
\caption{(a) Temperature dependence of the normalized critical current (full diamonds). 
The theoretical calculation (open circles) follows from the model of Ref. \cite{chrestin} with $Z=0.4$ (the model structure is sketched in the inset, see text). (b) Differential conductance vs. voltage for several temperature values. The weak but reproducible structure superimposed on the curves can be ascribed to multiple Andreev reflections (see text).
}
\label{F3}
\end{figure}


\begin{references}
\bibitem{etching}
	We employed a NH$_4$OH(25$\%$):DI water (1:5), 60 ml solution to etch the GaSb and AlSb layers, while for the InAs layer we used a CH$_3$COOH(99$\%$):H$_2$O$_2$(30$\%$):DI water  (1:5:1), 60 ml solution.
\bibitem{parametri} An effective electrode separation $L_{eff}=L+2L_t$ was employed in the fitting procedure. As pointed out in Ref. \cite{likharev} this is consistent with a coplanar geometry like the present one. The transfer length, i.e., the InAs  portion under each Nb contact actively participating to the transport, 
$L_t =280$ nm, was independently  measured by transfer length method (TLM) measurements on separate test structures.
\end{references}
\end{document}